# The Average Radial Speed of Light for The Rotating and Charged Black Hole

*Ting-Hang Pei

Thpei142857@gmail.com

**Abstract**-The Kerr-Newman metric is used to discuss the averagely measured speed of light along the radial direction at the black hole from a weak-gravitation reference frame such as an observer on Earth. The velocity equation of light at the black hole is represented in the spherical coordinate ($r$, $\theta$, $\phi$) and the main parameters are the Schwarzschild radius $R_S$, the rotation term $a$, and the charged term $R_Q$. From the calculations, the average radial speed of light from $r=R_S$ to $r=\alpha R_S$ with $\alpha > 1$ is possibly exceeding the speed of light $c$ in free space by an observer in a reference frame far away from the black hole like on Earth. The result extends to the large $r$ region when the rotation of the black hole is very high or the charge is large enough. This average radial speed finally goes to $c$ in a large distance away from the black hole. The results are reasonably at least for the radial directions from two poles and the place in the equatorial plane. We also propose a new explanation based on our results that the observation of the faster-than-light particle is due to the light bending near the black hole or supermassive star with very strong gravity. Finally, we give explanation that the propagation speed of gravity shall not be faster than the corresponding speed of light.

## 1. Introduction

The black hole has been studied more than one century. Its strong gravity attracts a lot of scientists to study the physics at the black hole. Traditional thoughts treat the black hole with a singularity collecting all mass and charges there. It is so mystery that all massive particles as well as the massless particles such as photons cannot escape its gravity. The non-rotational and uncharged black hole is defined by the Schwarzschild radius $R_S$ equal to $2GM/c^2$, where M is the mass of the black hole, $c$ is the speed of light in the free space, and G is the gravitational constant. According to General Relativity, its spacetime structure is tremendous changed and much different from the flat one such as the Minkowski spacetime structure. It causes the curiosity to think about what is the speed of light at the black hole measured by an observer in a reference frame like on Earth if possible?

Recently, the superluminal phenomena attract a lot of researchers and at the same time, some astronomical observations have been reported, especially the places close to the black holes [1-7]. So it also causes our curiosity to discuss these phenomena by using the Kerr-Newman metric [8-10] based on General Relativity. On one hand, the time dilation in astronomical observations have been observed many years [11-16], and the speed of light is indeed affected by gravity. According to the astronomical observations on Earth, the average speed of light is less than that in the free space. They provide a phenomenological proof that the averagely measured speed of light is changeable. On the other hand, the surface tangential speed of a rotating black hole has been found close to $c$ [17], so the discussion about the rotating effect on the speed of light is meaningful and to discuss the superluminal phenomenon becomes reasonable. In this research, the Kerr-Newman metric is used to discuss the averagely measured radial speed of light from the black hole and some special results are given.

## 2. The Kerr-Newman Metric and the Measured Velocity of Light at the Black Hole

When we discuss the propagation of light from the outer space through the event horizon into the black hole, the spacetime structure for the black hole is needed. There are three basic parameters to describe a black hole, the mass term $R_S$ with its total mass M, the rotation term $a$ with angular momentum $J$, and the charged term $R_Q$ with the total charge $Q$. There are several metrics to discuss the Einstein's spacetime structure, and the Kerr-Newman metric [8-10] is the one can simultaneously include these three parameters. Some other metrics [6-8] or alternative coordinates describing the spacetime structures for the black hole have been revealed many years. The expression for the Kerr-Newman metric in the polar coordinate ($t, r, \theta, \phi$) [18] is

$$ds^2 = -c^2 d\tau^2$$
$$= \left(\frac{dr^2}{\Delta} + d\theta^2\right)\rho^2 - (cdt - a\sin^2\theta d\phi)^2 \frac{\Delta}{\rho^2}$$
$$+ \left((r^2 + a^2)d\phi - acdt\right)^2 \frac{\sin^2\theta}{\rho^2}, \qquad (1)$$

where $\tau$ is the proper time, $t$ is the coordinate time,

$$\rho^2 = r^2 + a^2\cos^2\theta, \qquad (2)$$

$$\Delta = r^2 - rR_S + a^2 + R_Q^2, \qquad (3)$$

$a = J/Mc$, and $R_Q{}^2$ = $KQ^2G/c^4$ where K is the Coulomb's constant. The propagation of light is along the geodesic with $g_{uv}dx^u dx^v = ds^2 = 0$ [18-20], and it has been used to deduce the velocity of light in the Schwarzschild metric [19,20] and Kerr metric [18-20]. Then Eq. (1) gives an equation to describe three velocity components of light ($dr/dt$, $rd\theta/dt$,

$r\sin\theta d\phi/dt)$

$$\frac{\rho^4}{\Delta(\Delta-a^2\sin^2\theta)}\left(\frac{dr}{dt}\right)^2 + \frac{\rho^4}{r^2(\Delta-a^2\sin^2\theta)}\left(r\frac{d\theta}{dt}\right)^2$$
$$-\frac{(\Delta a^2\sin^2\theta - (r^2+a^2)^2)}{r^2(\Delta-a^2\sin^2\theta)}\left(r\sin\theta\frac{d\phi}{dt}\right)^2 - \frac{2ac(-\Delta+(r^2+a^2))\sin\theta}{r(\Delta-a^2\sin^2\theta)}\left(r\sin\theta\frac{d\phi}{dt}\right)$$
$$= c^2. \tag{4}$$

The relationship between each velocity component and the coordinate ($r$, $\theta$, $\phi$) is given in Eq. (4), and each velocity component must be real by observation. This way to deduce the light velocity has been used to deal with the in the Schwarzschild metric by an observer at infinity [18-20]. Because $ds^2 = -c^2 d\tau^2 = 0$ for light, calculating ($dr/d\tau$, $rd\theta/d\tau$, $r\sin\theta d\phi/d\tau$) is inappropriate for light. It is also pointed out that $\tau=0$ in the light reference frame [18]. Recently, the observations of the massive particles entering the black hole in the speed of $0.3c$ have been reported and it proves that the particle can fall into the black hole in very high speed [21]. In order to satisfy the truth that light can propagate into the black hole, the radial speed of light at the event horizon must be nonzero and the speeding time must be finite by the observations. However, the Kerr-Newman metric has a mathematical singularity at $r=0$ and $\theta = \pi/2$ due to a physical singularity with infinite mass density at the center. Such singularity is possibly removed. The finite-size nucleus in the black hole is a way to avoid this mathematical singularity.

### 3. The Conditions for using the Kerr-Newman Metric

According to Eq. (4), it permits one to calculate the speed of light at the black hole. A good start to discuss is choosing the geodesic of light only along the radial direction. Once an incident direction at certain $\theta$ is chosen, the velocity component $dr/dt$ of light can be only function of $r$. According to the equivalence principle in General Relativity, the time dilation is more explicitly as light is closer to the center of the black hole because of the stronger gravity. Using Eq. (4), then we can calculate how much time does light spend in or out of the black hole along the radial direction and its average speed from the measurement of an observer in a reference frame far away from the black hole like on earth. In such case, the geodesic is along the radial direction, and it gives the relation between $d\tau^2$ and $dt^2$, that is,

$$d\tau^2 = \frac{(\Delta - a^2\sin^2\theta)}{\rho^2} dt^2. \tag{5}$$

However, as we know, there are some singularities in the Kerr-Newman metric [10]. In order to really describe a physically reasonable black hole and avoid two event horizons [10], some conditions are required. Because we deal with a physical world and not pure mathematics, it has to describe the black hole more reasonably. The gravitational energy as well as the electric energy are both proportional to $1/r$, all mass and charges collected at the singularity at the center of the black seems to be very unreasonable. After all, the black hole is evolutional from the previous star that only has finitely total energy.

Then the transformation between the proper time and the time of the reference frame far away from the black hole in Eq. (5) is positive and it requires both denominator and numerator satisfying

$$\rho^2 > 0, \tag{6}$$

$$(\Delta - a^2 \sin^2\theta) > 0. \tag{7}$$

From Eq. (7), it can be expanded as

$$r^2 - rR_S + R_Q^2 + a^2\cos^2\theta > 0. \tag{8}$$

At $r=R_S/2$, Eq. (8) requires the condition

$$R_S^2 \leq 4(a^2\cos^2\theta + R_Q^2). \tag{9}$$

It is the condition at $r=R_S/2$ but not for other place $r>0$. Like at $r=R_S$, it only requires

$$R_Q^2 + a^2\cos^2\theta > 0, \tag{10}$$

and when $r>R_S$ Eq. (8) automatically exists. The other requirement is for the $dr^2$ term in Eq. (1) that is

$$\Delta > 0. \tag{11}$$

It also gives a condition at $r=R_S/2$

$$R_S^2 \leq 4(a^2 + R_Q^2). \tag{12}$$

However, similar to Eq. (10) at $r=R_S$, it only requires

$$R_Q^2 + a^2 > 0. \tag{13}$$

The results of Eqs. (9), (10), (12), and (13) seem to tell us the charged structure of a black hole. If we replace the concept of singularity with the finite-size nucleus in the black hole, then it can be explained and becomes reasonable. It means that the totally enclosed charges $Q$ is function of $r$ and has the expression $Q=Q(r)$ or

$$R_Q = R_Q(r). \tag{14}$$

Eq.(9) reveals that the totally enclosed charges have minimum requirement $R_Q > R_S/2$ at $r=R_S/2$ and Eqs. (10) and (13) imply that the region between $R_S/2$ can be occupied by the opposite charges so the totally enclosed charges at $r>R_S$ can be possibly very small even close to zero.

According to the equivalence principle in General Relativity, time dilation gives the other condition from Eq. (5) that

$$r \geq R_Q^2/R_S. \qquad (15)$$

As mentioned in Eq. (14), $R_Q$ is function of $r$, and $r$ continuously satisfies condition Eq. (15) from $r=0$, so the correct condition for any place between $r=0$ and $r=R_S$ is

$$r \geq R_Q^2/R_S \geq 0. \qquad (16)$$

This condition is physical and reasonable because the Kerr-Newman metric should correctly exist everywhere and not be bounded by some region or excluded by the singularity. If the time dilation is held correctly everywhere and no other physical mechanism limits this concept, then Eq. (16) gives the right condition. Otherwise, $r<R_Q^2/R_S$ would not be a well-defined space. At $r=R_S$, it further tells us that

$$R_S^2 \geq R_Q^2 \qquad (17)$$

## 4. The Average Speed of Light and the Superluminal Phenomenon In The Black Hole

After discussing the condition of $R_Q$, then we can calculate the spending time for light traveling from one place to another in or out of the black hole along the radial direction. The results are reasonably at least for the radial directions from two poles and the place in the equatorial plane. According to the principle of equivalence, in those place the accelerating directions are along the radial directions. Then from Eq. (4), the square velocity $(dr/dt)^2$ is

$$v_r^2 = \left(\frac{dr}{dr}\right)^2 = c^2\left[\frac{\Delta(\Delta - a^2\sin^2\theta)}{\rho^4}\right]. \qquad (18)$$

or

$$v_r^2 = \left(\frac{dr}{dr}\right)^2 = c^2\frac{(r^2 - rR_S + a^2 + R_Q^2)(r^2 - rR_S + a^2\cos^2\theta + R_Q^2)}{(r^2 + a^2\cos^2\theta)^2} \qquad (18')$$

Then it gives $dr/dt$

$$v_r = \frac{dr}{dt} = \pm c\frac{\sqrt{r^2 - rR_S + a^2 + R_Q^2}\sqrt{r^2 - rR_S + a^2\cos^2\theta + R_Q^2}}{r^2 + a^2\cos^2\theta} \qquad (19)$$

The sign '$\pm$' means that light can propagate forwardly and backwardly. It must be correct out of the black hole. Here we choose the positive (+) expression in Eq. (19) for convenient discussion. To calculate the spending time from $r=\alpha_1 R_S$ to $r=\alpha_2 R_S$ for non-negative $\alpha_1$ and $\alpha_2$ and different $\theta$, it needs to integrate the time $t$ and the radial distance $r$ by moving $dr$ and $dt$ to different side of Eq. (19). The spending time $T_{Measure}$ is the time measured by an observer in a reference frame like on Earth or the center of the black hole where the gravitation is zero for a black hole with finite-size nucleus. Then the integrations are

$$\int_{\alpha_1 R_S}^{\alpha_2 R_S} \frac{r^2 + a^2\cos^2\theta}{\sqrt{r^2 - rR_S + a^2 + R_Q^2}\sqrt{r^2 - rR_S + a^2\cos^2\theta + R_Q^2}} dr = c\int_0^{T_{measure}} dt. \quad (20)$$

Setting

$$y = r^2 - rR_S + R_Q^2 = \left(r - \frac{R_S}{2}\right)^2 - \frac{R_S^2}{4} + R_Q^2, \quad (21)$$

it gives

$$y \geq -\frac{R_S^2}{4} + R_Q^2 \quad (22)$$

and $y>0$ when we consider $r>R_S$. Then the expression of $r$ is

$$r = \frac{R_S \pm 2\sqrt{y + \frac{R_S^2}{4} - R_Q^2}}{2}. \quad (23)$$

Substituting Eq. (21) into Eq. (23), the square-root term is non-imaginary, and + is for $r>R_S/2$ and – is for $r \leq R_S/2$ in Eq. (23). From this expression then we have

$$dy = 2\left(r - \frac{R_S}{2}\right)dr, \quad (24)$$

or

$$dr = \pm \frac{dy}{2\sqrt{y + \frac{R_S^2}{4} - R_Q^2}}. \quad (24')$$

Substituting Eq. (24') into Eq. (20) and considering + for $r>R_S/2$ and – for $r \leq R_S/2$, then we have the measurement time $T_{measure}$

$$cT_{measure} = \int_{(\alpha_1^2 - \alpha_1)R_S^2 + R_Q^2}^{(\alpha_2^2 - \alpha_2)R_S^2 + R_Q^2} \frac{\left(y + \frac{R_S^2}{2} - R_Q^2 + R_S\sqrt{y + \frac{R_S^2}{4} - R_Q^2} + a^2\cos^2\theta\right)}{2\sqrt{y + a^2}\sqrt{y + a^2\cos^2\theta}\sqrt{y + \frac{R_S^2}{4} - R_Q^2}} dy$$

$$= \int_{(\alpha_1^2 - \alpha_1)R_S^2 + R_Q^2}^{(\alpha_2^2 - \alpha_2)R_S^2 + R_Q^2} \frac{\sqrt{y + \frac{R_S^2}{4} - R_Q^2}}{2\sqrt{y + a^2}\sqrt{y + a^2\cos^2\theta}} dy$$

$$+\left\{\times \int_{(\alpha_1^2-\alpha_1)R_S^2+R_Q^2}^{(\alpha_2^2-\alpha_2)R_S^2+R_Q^2}\frac{\left(\frac{R_S^2}{4}+a^2\cos^2\theta\right)}{2\sqrt{y+a^2}\sqrt{y+a^2\cos^2\theta}\sqrt{y+\frac{R_S^2}{4}-R_Q^2}}dy\right\}$$

$$+\int_{(\alpha_1^2-\alpha_1)R_S^2+R_Q^2}^{(\alpha_2^2-\alpha_2)R_S^2+R_Q^2}\frac{R_S}{2\sqrt{y+a^2}\sqrt{y+a^2\cos^2\theta}}dy$$

$$=I_1+I_2+I_3. \tag{25}$$

These three parts of the total Integral, $I_1$, $I_2$, $I_3$, show the different forms of the elliptical Integrals. Then we set $\{\zeta,\omega,\eta\}=\{(-a^2),(-a^2\cos^2\theta),(-R_S^2/4+R_Q^2)\}$ with the condition $\zeta>\omega>\eta$ in integrals. Without considering $\cos\theta=1$, it gives three possible situations and corresponding parameters $\mu_1$, $\mu_2$, and $q$ [22].

**Situation 1** $(\zeta=-R_S^2/4+R_Q^2,\ \omega=-a^2\cos^2\theta,\eta=-a^2)$:

$$(-R_S^2/4+R_Q^2)>(-a^2\cos^2\theta)>(-a^2), \tag{26}$$

and

$$\mu_1=\arcsin\sqrt{\frac{\left(\alpha_1^2-\alpha_1+\frac{1}{4}\right)R_S^2}{[(\alpha_1^2-\alpha_1)R_S^2+R_Q^2+a^2\cos^2\theta]}}, \tag{27a}$$

$$\mu_2=\arcsin\sqrt{\frac{\left(\alpha_2^2-\alpha_2+\frac{1}{4}\right)R_S^2}{[(\alpha_2^2-\alpha_2)R_S^2+R_Q^2+a^2\cos^2\theta]}}, \tag{27b}$$

$$q=\sqrt{\frac{a^2(1-\cos^2\theta)}{\left(-\frac{R_S^2}{4}+R_Q^2+a^2\right)}}. \tag{27c}$$

**Situation 2** $(\zeta=-a^2\cos^2\theta,\omega=-R_S^2/4+R_Q^2,\eta=-a^2)$:

$$(-a^2\cos^2\theta)>(-R_S^2/4+R_Q^2)>(-a^2), \tag{28}$$

and

$$\mu_1=\arcsin\sqrt{\frac{[(\alpha_1^2-\alpha_1)R_S^2+R_Q^2+a^2\cos^2\theta]}{\left(\alpha_1^2-\alpha_1+\frac{1}{4}\right)R_S^2}}, \tag{29a}$$

$$\mu_2 = \arcsin \sqrt{\frac{[(\alpha_2^2 - \alpha_2)R_S^2 + R_Q^2 + a^2\cos^2\theta]}{\left(\alpha_2^2 - \alpha_2 + \frac{1}{4}\right)R_S^2}}, \quad (29b)$$

$$q = \sqrt{\frac{\left(-\frac{R_S^2}{4} + R_Q^2 + a^2\right)}{a^2(1 - \cos^2\theta)}}. \quad (29c)$$

**Situation 3** $\left(\zeta = -a^2\cos^2\theta, \omega = -a^2, \eta = -R_S^2/4 + R_Q^2\right)$:

$$(-a^2\cos^2\theta) > (-a^2) > \left(-R_S^2/4 + R_Q^2\right), \quad (30)$$

and

$$\mu_1 = \arcsin \sqrt{\frac{[(\alpha_1^2 - \alpha_1)R_S^2 + R_Q^2 + a^2\cos^2\theta]}{[(\alpha_1^2 - \alpha_1)R_S^2 + R_Q^2 + a^2]}}, \quad (31a)$$

$$\mu_2 = \arcsin \sqrt{\frac{[(\alpha_2^2 - \alpha_2)R_S^2 + R_Q^2 + a^2\cos^2\theta]}{[(\alpha_2^2 - \alpha_2)R_S^2 + R_Q^2 + a^2]}}, \quad (31b)$$

$$q = \sqrt{\frac{\left(\frac{R_S^2}{4} - R_Q^2 - a^2\right)}{\left(\frac{R_S^2}{4} - R_Q^2 - a^2\cos^2\theta\right)}}. \quad (31c)$$

Supposing $y_{\text{lower}} = \max\left\{(-a^2), (-a^2\cos^2\theta), \left(-R_S^2/4 + R_Q^2\right)\right\}$, then the first two integrals become

$$I_1 = \int_{y_{lower}}^{(\alpha_2^2 - \alpha_2)R_S^2 + R_Q^2} \frac{\sqrt{y - \left(-\frac{R_S^2}{4} + R_Q^2\right)}}{2\sqrt{y - (-a^2)}\sqrt{y - (-a^2\cos^2\theta)}} dy$$

$$- \int_{y_{lower}}^{(\alpha_1^2 - \alpha_1)R_S^2 + R_Q^2} \frac{\sqrt{y - \left(-\frac{R_S^2}{4} + R_Q^2\right)}}{2\sqrt{y - (-a^2)}\sqrt{y - (-a^2\cos^2\theta)}} dy \quad (32)$$

$$I_2 = \left(\frac{R_S^2}{4} + a^2\cos^2\theta\right) \times$$

$$\int_{y_{lower}}^{(\alpha_2^2-\alpha_2)R_S^2+R_Q^2} \frac{1}{2\sqrt{y-(-a^2)}\sqrt{y-(-a^2\cos^2\theta)}\sqrt{y-\left(-\frac{R_S^2}{4}+R_Q^2\right)}} dy$$

$$-\left(\frac{R_S^2}{4}+a^2\cos^2\theta\right) \times$$

$$\int_{y_{lower}}^{(\alpha_1^2-\alpha_1)R_S^2+R_Q^2} \frac{1}{2\sqrt{y-(-a^2)}\sqrt{y-(-a^2\cos^2\theta)}\sqrt{y-\left(-\frac{R_S^2}{4}+R_Q^2\right)}} dy. \quad (33)$$

The third integral has the solution as

$$I_3 = \frac{R_S}{2} \ln\left[2\sqrt{y^2+a^2(1+\cos^2\theta)y+a^4\cos^2\theta}\right.$$

$$\left. + 2y + a^2(1+\cos^2\theta)\right]\Big|_{(\alpha_1^2-\alpha_1)R_S^2+R_Q^2}^{(\alpha_2^2-\alpha_2)R_S^2+R_Q^2} \quad (34)$$

According to these three situations, the total Integrals are also divided into three results accompanying three parameters $\mu_1, \mu_2$, and $q$. These results are suitable for

$$(\alpha_2^2-\alpha_2)R_S^2 + R_Q^2 > (\alpha_1^2-\alpha_1)R_S^2 + R_Q^2 > y_{lower}. \quad (35)$$

However, it automatically exists for any non-negative $\alpha_1$ and $\alpha_2$. When $\alpha_1=0$, Eq. (35) gives $R_Q^2 > y_{lower}$ and Situations 2 and 3 satisfy it because $y_{lower} < 0$ and $R_Q^2 \geq 0$. When we check Situation 1, it gives

$$y_{lower} = -\frac{R_S^2}{4} + R_Q^2 < R_Q^2. \quad (36)$$

It is based on the factor that $R_S^2 > 0$. Then these three results are list as follows.

**Results 1** for Situation 1:

$$I_1 + I_2 + I_3 = -2\sqrt{\left(-\frac{R_S^2}{4}+R_Q^2\right)+a^2}\left(E(\mu_2,q)-E(\mu_1,q)\right)$$

$$+2\sqrt{\frac{\left(\alpha_2^2-\alpha_2+\frac{1}{4}\right)R_S^2\left[(\alpha_2^2-\alpha_2)R_S^2+R_Q^2+a^2\right]}{\left[(\alpha_2^2-\alpha_2)R_S^2+R_Q^2+a^2\cos^2\theta\right]}}$$

$$-2\sqrt{\frac{\left(\alpha_1^2 - \alpha_1 + \frac{1}{4}\right) R_S^2 [(\alpha_1^2 - \alpha_1)R_S^2 + R_Q^2 + a^2]}{[(\alpha_1^2 - \alpha_1)R_S^2 + R_Q^2 + a^2\cos^2\theta]}}$$

$$+\left(\frac{R_S^2}{4} + a^2\cos^2\theta\right)\frac{1}{\sqrt{\left(-\frac{R_S^2}{4} + R_Q^2\right) + a^2}}(F(\mu_2, q) - F(\mu_1, q)) +$$

$$\frac{R_S}{2}\ln\left\{2\sqrt{[(\alpha_2^2 - \alpha_2)R_S^2 + R_Q^2]^2 + a^2(1+\cos^2\theta)[(\alpha_2^2 - \alpha_2)R_S^2 + R_Q^2] + a^4\cos^2\theta}\right.$$

$$\left. + 2[(\alpha_2^2 - \alpha_2)R_S^2 + R_Q^2] + a^2(1+\cos^2\theta)\right\} -$$

$$\frac{R_S}{2}\ln\left\{2\sqrt{[(\alpha_1^2 - \alpha_1)R_S^2 + R_Q^2]^2 + a^2(1+\cos^2\theta)[(\alpha_1^2 - \alpha_1)R_S^2 + R_Q^2] + a^4\cos^2\theta}\right.$$

$$\left. + 2[(\alpha_1^2 - \alpha_1)R_S^2 + R_Q^2] + a^2(1+\cos^2\theta)\right\}, \tag{37}$$

where $E(\mu, q)$ and $F(\mu, q)$ are the incomplete elliptical integrals of the first and second kinds, respectively.

**Result 2** for Situation 2:

$$I_1 + I_2 + I_3 = \frac{2\left(a^2\cos^2\theta + \frac{R_S^2}{4} - R_Q^2\right)}{\sqrt{a^2(1 - \cos^2\theta)}}(F(\mu_2, q) - F(\mu_1, q))$$

$$-2\sqrt{a^2(1 - \cos^2\theta)}(E(\mu_2, q) - E(\mu_1, q))$$

$$+2\sqrt{\frac{[(\alpha_2^2 - \alpha_2)R_S^2 + R_Q^2 + a^2\cos^2\theta][(\alpha_2^2 - \alpha_2)R_S^2 + R_Q^2 + a^2]}{\left(\alpha_2^2 - \alpha_2 + \frac{1}{4}\right)R_S^2}}$$

$$-2\sqrt{\frac{[(\alpha_1^2 - \alpha_1)R_S^2 + R_Q^2 + a^2\cos^2\theta][(\alpha_1^2 - \alpha_1)R_S^2 + R_Q^2 + a^2]}{\left(\alpha_1^2 - \alpha_1 + \frac{1}{4}\right)R_S^2}}$$

$$+\left(\frac{R_S^2}{4} + a^2\cos^2\theta\right)\frac{1}{\sqrt{a^2(1-\cos^2\theta)}}(F(\mu_2, q) - F(\mu_1, q)) +$$

$$\frac{R_S}{2} \ln\left\{2\sqrt{[(\alpha_2^2 - \alpha_2)R_S^2 + R_Q^2]^2 + a^2(1+\cos^2\theta)[(\alpha_2^2 - \alpha_2)R_S^2 + R_Q^2] + a^4\cos^2\theta}\right.$$

$$\left. + 2[(\alpha_2^2 - \alpha_2)R_S^2 + R_Q^2] + a^2(1+\cos^2\theta)\right\} -$$

$$\frac{R_S}{2} \ln\left\{2\sqrt{[(\alpha_1^2 - \alpha_1)R_S^2 + R_Q^2]^2 + a^2(1+\cos^2\theta)[(\alpha_1^2 - \alpha_1)R_S^2 + R_Q^2] + a^4\cos^2\theta}\right.$$

$$\left. + 2[(\alpha_1^2 - \alpha_1)R_S^2 + R_Q^2] + a^2(1+\cos^2\theta)\right\}. \tag{38}$$

**Result 3** for Situation 3:

$$I_1 + I_2 + I_3 =$$

$$2\sqrt{a^2\cos^2\theta - \left(-\frac{R_S^2}{4} + R_Q^2\right)}\left[(F(\mu_2,q) - F(\mu_1,q)) - (E(\mu_2,q) - E(\mu_1,q))\right]$$

$$+2\sqrt{\frac{[(\alpha_2^2 - \alpha_2)R_S^2 + R_Q^2 + a^2\cos^2\theta]\left(\alpha_2^2 - \alpha_2 + \frac{1}{4}\right)R_S^2}{[(\alpha_2^2 - \alpha_2)R_S^2 + R_Q^2 + a^2]}}$$

$$-2\sqrt{\frac{[(\alpha_1^2 - \alpha_1)R_S^2 + R_Q^2 + a^2\cos^2\theta]\left(\alpha_1^2 - \alpha_1 + \frac{1}{4}\right)R_S^2}{[(\alpha_1^2 - \alpha_1)R_S^2 + R_Q^2 + a^2]}}$$

$$+\left(\frac{R_S^2}{4} + a^2\cos^2\theta\right)\frac{1}{\sqrt{a^2\cos^2\theta - \left(-\frac{R_S^2}{4} + R_Q^2\right)}}(F(\mu_2,q) - F(\mu_1,q)) +$$

$$\frac{R_S}{2} \ln\left\{2\sqrt{[(\alpha_2^2 - \alpha_2)R_S^2 + R_Q^2]^2 + a^2(1+\cos^2\theta)[(\alpha_2^2 - \alpha_2)R_S^2 + R_Q^2] + a^4\cos^2\theta}\right.$$

$$\left. + 2[(\alpha_2^2 - \alpha_2)R_S^2 + R_Q^2] + a^2(1+\cos^2\theta)\right\} -$$

$$\frac{R_S}{2} \ln\left\{2\sqrt{[(\alpha_1^2 - \alpha_1)R_S^2 + R_Q^2]^2 + a^2(1+\cos^2\theta)[(\alpha_1^2 - \alpha_1)R_S^2 + R_Q^2] + a^4\cos^2\theta}\right.$$

$$\left. + 2[(\alpha_1^2 - \alpha_1)R_S^2 + R_Q^2] + a^2(1+\cos^2\theta)\right\}. \tag{39}$$

Results 1 to 3 mean the measurements of the average radial speed of light at different order of $\{\zeta, \omega, \eta\}$ from a reference frame far away from the black hole like on Earth. Because we only consider the inequalities between $\zeta$, $\omega$, and $\eta$, we have to further consider two of them or all of them that are equal. Considering $\eta=\zeta$, we have $R_S^2/4 = R_Q^2 + a^2$ in Eq. (25), the integral $I_1+I_2+I_3$ for $|\cos\theta| < 1$ is

$$\int_{(\alpha_1^2-\alpha_2)R_S^2+R_Q^2}^{(\alpha_2^2-\alpha_2)R_S^2+R_Q^2} \frac{1}{2\sqrt{y+a^2\cos^2\theta}} dy +$$

$$\left(\frac{R_S^2}{4} + a^2\cos^2\theta\right) \int_{(\alpha_1^2-\alpha_1)R_S^2+R_Q^2}^{(\alpha_2^2-\alpha_2)R_S^2+R_Q^2} \frac{1}{2(y+a^2)\sqrt{y+a^2\cos^2\theta}} dy +$$

$$\int_{(\alpha_1^2-\alpha_1)R_S^2+R_Q^2}^{(\alpha_2^2-\alpha_2)R_S^2+R_Q^2} \frac{R_S}{2\sqrt{y+a^2}\sqrt{y+a^2\cos^2\theta}} dy$$

$$= \sqrt{y+a^2\cos^2\theta}\Big|_{(\alpha_1^2-\alpha_1)R_S^2+R_Q^2}^{(\alpha_2^2-\alpha_2)R_S^2+R_Q^2} +$$

$$\left(\frac{R_S^2}{4} + a^2\cos^2\theta\right)\frac{1}{a\sin\theta}\tan^{-1}\sqrt{\frac{y+a^2\cos^2\theta}{a^2\sin^2\theta}}\Bigg|_{(\alpha_1^2-\alpha_1)R_S^2+R_Q^2}^{(\alpha_2^2-\alpha_2)R_S^2+R_Q^2} +$$

$$\frac{R_S}{2}\ln\left[2\sqrt{y^2+a^2(1+\cos^2\theta)y+a^4\cos^2\theta}+2y+a^2(1+\cos^2\theta)\right]\Bigg|_{(\alpha_1^2-\alpha_1)R_S^2+R_Q^2}^{(\alpha_2^2-\alpha_2)R_S^2+R_Q^2}$$

$$= \sqrt{(\alpha_2^2-\alpha_2)R_S^2+R_Q^2+a^2\cos^2\theta} - \sqrt{(\alpha_1^2-\alpha_1)R_S^2+R_Q^2+a^2\cos^2\theta} +$$

$$\left(\frac{R_S^2}{4}+a^2\cos^2\theta\right)\frac{1}{a\sin\theta}\left[\tan^{-1}\sqrt{\frac{[(\alpha_2^2-\alpha_2)R_S^2+R_Q^2+a^2\cos^2\theta]}{a^2\sin^2\theta}}\right.$$

$$\left. - \tan^{-1}\sqrt{\frac{[(\alpha_1^2-\alpha_1)R_S^2+R_Q^2+a^2\cos^2\theta]}{a^2\sin^2\theta}}\right] +$$

$$\frac{R_S}{2}\ln\left\{2\sqrt{[(\alpha_2^2-\alpha_2)R_S^2+R_Q^2]^2+a^2(1+\cos^2\theta)[(\alpha_2^2-\alpha_2)R_S^2+R_Q^2]+a^4\cos^2\theta}\right.$$

$$\left. + 2[(\alpha_2^2-\alpha_2)R_S^2+R_Q^2]+a^2(1+\cos^2\theta)\right\} -$$

$$\frac{R_S}{2}\ln\left\{2\sqrt{[(\alpha_1^2-\alpha_1)R_S^2+R_Q^2]^2+a^2(1+\cos^2\theta)[(\alpha_1^2-\alpha_1)R_S^2+R_Q^2]+a^4\cos^2\theta}\right.$$

$$\left. + 2[(\alpha_1^2-\alpha_1)R_S^2+R_Q^2]+a^2(1+\cos^2\theta)\right\}. \tag{40}$$

Then considering $\eta=\omega$, we have $R_S^2/4 = R_Q^2 + a^2\cos^2\theta$ in Eq. (25), the integral $I_1+I_2+I_3$ for $|\cos\theta| < 1$ is

$$\int_{(\alpha_1^2-\alpha_2)R_S^2+R_Q^2}^{(\alpha_2^2-\alpha_2)R_S^2+R_Q^2} \frac{1}{2\sqrt{y+a^2}} dy +$$

$$\left(\frac{R_S^2}{4}+a^2\cos^2\theta\right)\int_{(\alpha_1^2-\alpha_1)R_S^2+R_Q^2}^{(\alpha_2^2-\alpha_2)R_S^2+R_Q^2} \frac{1}{2(y+a^2\cos^2\theta)\sqrt{y+a^2}} dy +$$

$$\int_{(\alpha_1^2-\alpha_1)R_S^2+R_Q^2}^{(\alpha_2^2-\alpha_2)R_S^2+R_Q^2} \frac{R_S}{2\sqrt{y+a^2}\sqrt{y+a^2\cos^2\theta}} dy$$

$$= \sqrt{y+a^2}\Big|_{(\alpha_1^2-\alpha_1)R_S^2+R_Q^2}^{(\alpha_2^2-\alpha_2)R_S^2+R_Q^2} +$$

$$\left(\frac{R_S^2}{4}+a^2\cos^2\theta\right)\frac{1}{a\sin\theta}\ln\left(\frac{\sqrt{y+a^2}-a\sin\theta}{\sqrt{y+a^2}+a\sin\theta}\right)\Big|_{(\alpha_1^2-\alpha_1)R_S^2+R_Q^2}^{(\alpha_2^2-\alpha_2)R_S^2+R_Q^2} +$$

$$\frac{R_S}{2}\ln\left[2\sqrt{y^2+a^2(1+\cos^2\theta)y+a^4\cos^2\theta}+2y+a^2(1+\cos^2\theta)\right]\Big|_{(\alpha_1^2-\alpha_1)R_S^2+R_Q^2}^{(\alpha_2^2-\alpha_2)R_S^2+R_Q^2}$$

$$= \sqrt{(\alpha_2^2-\alpha_2)R_S^2+R_Q^2+a^2} - \sqrt{(\alpha_1^2-\alpha_1)R_S^2+R_Q^2+a^2} +$$

$$\left(\frac{R_S^2}{4}+a^2\cos^2\theta\right)\frac{1}{a\sin\theta}\left[\ln\left(\frac{\sqrt{(\alpha_2^2-\alpha_2)R_S^2+R_Q^2+a^2}-a\sin\theta}{\sqrt{(\alpha_2^2-\alpha_2)R_S^2+R_Q^2+a^2}+a\sin\theta}\right)\right.$$

$$\left. - \ln\left(\frac{\sqrt{(\alpha_1^2-\alpha_1)R_S^2+R_Q^2+a^2}-a\sin\theta}{\sqrt{(\alpha_1^2-\alpha_1)R_S^2+R_Q^2+a^2}+a\sin\theta}\right)\right] +$$

$$\frac{R_S}{2}\ln\left\{2\sqrt{[(\alpha_2^2-\alpha_2)R_S^2+R_Q^2]^2+a^2(1+\cos^2\theta)[(\alpha_2^2-\alpha_2)R_S^2+R_Q^2]+a^4\cos^2\theta}\right.$$

$$\left. + 2[(\alpha_2^2-\alpha_2)R_S^2+R_Q^2]+a^2(1+\cos^2\theta)\right\} +$$

$$\frac{R_S}{2}\ln\left\{2\sqrt{[(\alpha_1^2-\alpha_1)R_S^2+R_Q^2]^2+a^2(1+\cos^2\theta)[(\alpha_1^2-\alpha_1)R_S^2+R_Q^2]+a^4\cos^2\theta}\right.$$

$$\left. + 2[(\alpha_1^2-\alpha_1)R_S^2+R_Q^2]+a^2(1+\cos^2\theta)\right\}. \tag{41}$$

Considering $\zeta = \omega \neq \eta$, we have the case at two poles: $|\cos\theta| = 1$, but $R_S^2/4 \neq R_Q^2+a^2$. The speed of the integral in Eq. (20) can be directly calculated. Then one can proceed the integral

$$L = \int_{\alpha_1 R_S}^{\alpha_2 R_S} \left[ 1 - \frac{R_Q^2}{r^2 - rR_S + a^2 + R_Q^2} + \frac{rR_S}{r^2 - rR_S + a^2 + R_Q^2} \right] dr. \qquad (42)$$

For the case of $4(a^2 + R_Q^2) > R_S^2$, it gives

$$L = (\alpha_2 - \alpha_1)R_S$$

$$- \left( R_Q^2 - \frac{R_S^2}{2} \right) \left( \frac{2}{\sqrt{4(a^2 + R_Q^2) - R_S^2}} \right) \left( \tan^{-1} \frac{2r - R_S}{\sqrt{4(a^2 + R_Q^2) - R_S^2}} \right) \Bigg|_{\alpha_1 R_S}^{\alpha_2 R_S}$$

$$+ \frac{R_S}{2} \ln |r^2 - rR_S + a^2 + R_Q^2| \Bigg|_{\alpha_1 R_S}^{\alpha_2 R_S}$$

$$= (\alpha_2 - \alpha_1)R_S$$

$$- \frac{(2R_Q^2 - R_S^2)}{\sqrt{4(a^2 + R_Q^2) - R_S^2}} \left( \tan^{-1} \frac{(2\alpha_2 - 1)R_S}{\sqrt{4(a^2 + R_Q^2) - R_S^2}} - \tan^{-1} \frac{(2\alpha_1 - 1)R_S}{\sqrt{4(a^2 + R_Q^2) - R_S^2}} \right)$$

$$+ \frac{R_S}{2} \ln \frac{|(\alpha_2^2 - \alpha_2)R_S^2 + a^2 + R_Q^2|}{|(\alpha_1^2 - \alpha_1)R_S^2 + a^2 + R_Q^2|}$$

$$= cT_{Measure}. \qquad (43)$$

For the case of $4(a^2 + R_Q^2) < R_S^2$, it gives

$$L = (\alpha_2 - \alpha_1)R_S$$

$$- \frac{R_Q^2 - \frac{R_S^2}{2}}{\sqrt{R_S^2 - 4(a^2 + R_Q^2)}} \left( \ln \frac{2r - R_S - \sqrt{R_S^2 - 4(a^2 + R_Q^2)}}{2r - R_S + \sqrt{R_S^2 - 4(a^2 + R_Q^2)}} \right) \Bigg|_{\alpha_1 R_S}^{\alpha_2 R_S}$$

$$+ \frac{R_S}{2} \ln |r^2 - rR_S + a^2 + R_Q^2| \Bigg|_{\alpha_1 R_S}^{\alpha_2 R_S}$$

$$= (\alpha_2 - \alpha_1)R_S - \frac{R_Q^2 - \frac{R_S^2}{2}}{\sqrt{R_S^2 - 4(a^2 + R_Q^2)}} \times$$

$$\ln \left( \frac{(2\alpha_2 - 1)R_S - \sqrt{R_S^2 - 4(a^2 + R_Q^2)}}{(2\alpha_2 - 1)R_S + \sqrt{R_S^2 - 4(a^2 + R_Q^2)}} \cdot \frac{(2\alpha_1 - 1)R_S + \sqrt{R_S^2 - 4(a^2 + R_Q^2)}}{(2\alpha_1 - 1)R_S - \sqrt{R_S^2 - 4(a^2 + R_Q^2)}} \right)$$

$$+ \frac{R_S}{2} \ln \frac{|(\alpha_2^2 - \alpha_2)R_S^2 + a^2 + R_Q^2|}{|(\alpha_1^2 - \alpha_1)R_S^2 + a^2 + R_Q^2|}.$$

$$= cT_{Measure}. \tag{44}$$

For the case of $a^2 = a^2\cos^2\theta$ and $4(a^2 + R_Q^2) = R_S^2$, it gives

$$L = (\alpha_2 - \alpha_1)R_S + \left(R_Q^2 - \frac{R_S^2}{2}\right)\left(\frac{2}{2r - R_S}\right)\Big|_{\alpha_1 R_S}^{\alpha_2 R_S} + R_S \ln\left(r - \frac{R_S}{2}\right)\Big|_{\alpha_1 R_S}^{\alpha_2 R_S}$$

$$= (\alpha_2 - \alpha_1)R_S + \frac{2R_Q^2 - R_S^2}{R_S}\left(\frac{1}{2\alpha_2 - 1} - \frac{1}{2\alpha_1 - 1}\right) + R_S \ln\frac{2\alpha_2 - 1}{2\alpha_1 - 1}.$$

$$= cT_{Measure}. \tag{45}$$

Substituting Eqs. (37) - (41) and (43) – (45) into Eq. (25), it gives the measurement time

$$T_{measrure} = \frac{1}{c}(I_1 + I_2 + I_3), \tag{46}$$

which has different value dependent on $a$, $R_Q$, and $\theta$. The distance is $(\alpha_2-\alpha_1)R_S$ and the measurement time $T_{Earth}$ for light traveling this distance on Earth should be about

$$T_{Earth} = \frac{1}{c}(\alpha_2 - \alpha_1)R_S. \tag{47}$$

$T_{Measure}$ is possible less than $T_{Earth}$. According to this, the average speed of light $c_{ave}$ from $r=\alpha_1 R_S$ to $r=\alpha_2 R_S$ is

$$c_{ave} = c(\alpha_2 - \alpha_1)R_s/(I_1 + I_2 + I_3). \tag{48}$$

So we can calculate the average radial speed of light $c_{ave}$. It is easy to prove that when $\alpha_2 \gg \alpha_1$ and $\alpha_2 \gg 1$, Eq.(48) gives

$$c_{ave} \sim c. \tag{49}$$

This result is the reasonable speed of light that we measure on Earth.

## 5. Demonstrated Cases of the Superluminal Phenomena outer of the Black Hole

Next, we discuss the possibility whether the superluminal phenomenon can take place at the place larger than $R_S$. In astronomical observations, the measurements are during a finite period. Let $\alpha_1=1$ for calculating the average radial speed of light on the two-dimensional x-y plane with the y-axis parallel to the rotational axis passing through the center of the black hole, the integrals in Eq. (20) or (25) can be calculated numerically for different $a$, $R_Q$, and $\alpha_2$ cases. The center of the black hole is set as the origin of the coordinate. In order to satisfy the condition in Eq. (17), we consider the maximum of $R_Q$ is $R_S$ for $r≥R_S$.

In order to understand the effect of rotational term $a$ with the small $R_Q=0.1R_S$, first we study two cases of $a=R_S$ and $a=100R_S$. In those cases maximal x and y values are 10, 100, and 200$R_S$ as shown in Figs. 1(a) – (c). The calculating space is divided into 1000x1000 square grids, and the date in each grid represents the average radial speed in unit of $c$ calculated from $r=R_S$ to that point so all the data $r \leq R_S$ are zero in figures. The grid size is $(0.02R_S)^2$ for the calculation of the maximal x=y=10$R_S$ case, $(0.2R_S)^2$ for the calculation of the maximal x=y=100$R_S$ case, and $(0.4R_S)^2$ for the calculation of the maximal x=y=200$R_S$ case. These parameters are the same in all corresponding Figs. 1 – 6. In Fig. 1(a), the distribution of the average radial speed of light for $a=R_S$ shows that all average radial speeds are less than $c$, and it becomes slower and slower as $r$ gradually closes $R_S$ and then gradually increases as $r$ leaves away $R_S$. At the same distance away from the center, the average radial speed is the slowest one on the x-axis and the highest one on the y-axis. In Figs. 1(b) and (c), the distributions show the average radial speed in the larger range and both of them also reveal the increase trends as $r$ is away from the center. The data along the x-axis in Fig. 1(c) is draw in Fig. 1(d) where the maximum of $r$ is 200$R_S$ symmetrical to the center. It shows that the average radial speed of light is only about $0.2c$ at the place adjacent to $R_S$ and gradually close to $c$ far away from the center.

Then we consider the case of $a=100R_S$ and $R_Q=0.1R_S$. It is much explicit that the highly average radial speed more than $20.0c$ mainly distributes in a very narrow range near the x-axis as shown in Fig. 2(a) and the maximum is close to $45.0c$. The average radial speed quickly drops to about $10c$ as $r=20R_S$ as shown in Figs. 2(b) and (c). As $r$ increases, it gradually goes to $c$. Comparing the calculations with those in Fig. 1, the highest average speed is not on the y-axis. On the contrary, the highest value appears on the x-axis and the slowest one on the y-axis at the same $r$. Similar to Fig. 1(d), the distribution along the x-axis is shown in Fig. 2(d) where the maximum of $r$ is 200$R_S$ symmetrical to the center. Specially, the maximum on the x-axis is not at the place adjacent to $R_S$ but a little away from $R_S$. The maximum is at about $r=2R_S$.

The third case is similar to the first one that $a=R_S$ but $R_Q$ increases from 0.1 $R_S$ to 1.0$R_S$ However, the results are much different from those in Fig. 1. In Fig. 3(a), the maximum is at the place adjacent to $R_S$ on the x-axis and a lot of places show the average radial speeds more than $0.8c$. So increasing $R_Q$ from $0.1R_S$ to $1.0R_S$ also raising the average radial speed and changes the distribution. In Figs. 3(b) and (c), the distributions in the larger ranges show that the variation along the x-axis is larger than that along the y-axis and the equi-speed surfaces are elliptical shapes. The distribution along the x-axis in Fig. 3(c) is drawn in Fig. 3(d) where the average radial speed is about $1.4c$ adjacent to $R_S$ and quickly drops to the minimum roughly $0.85c$ at about $r=4.0R_S$. The maximum of $r$ is 200$R_S$ symmetrical to the center in Fig. 3(d). After this minimum, it slowly increases to close to $1.0c$. Along each direction, all average radial speeds show the similar trends reaching $1.0c$ when $r$ increases largely.

The fourth case holds $R_Q$ equal to $1.0R_S$ but increases $a$ to $10.0R_S$. In Fig. 4(a), the maximum about $10.0c$ is also adjacent to $R_S$ on the x-axis. The average radial speed more than half maximum distributes in a narrow range near the x-axis but is not like the result in Fig. 2 that the distribution is more centralized to the x-axis. As enlarging the calculation range, the average radial speed more than $2.0c$ mainly distributes near the x-axis and very close to $R_S$ as shown in Figs. 4(b) and (c). The distribution of the average radial speed along the x-axis is drawn in Fig. 4(d), in which it drops very quickly from the place adjacent to $R_S$ to $200R_S$. Finally, the average radial speed is close to $1.0c$ as $r$ is very large in our calculations.

The fifth case uses the same $R_Q=1.0R_S$ and $a$ is increased to $50R_S$ in order to investigate the effect of the rotation on the average radial speed. In Fig. 5(a), the distribution of the average radial speed more than $20.0c$ is very close to the axis symmetric to the center within $4.0R_S$. The maximum about $50.0c$ is also adjacent to $R_S$ on the x-axis. Comparing Fig. 5(a) with 4(a), it shows this distribution closer to the x-axis and more centralized. In the enlarged space as shown in Figs. 5(b) and (c), the average radial speed more than $20.0c$ is close to the center. The distribution of the average radial speed along the x-axis is shown in Fig. 5(d) where it quickly drops from the place adjacent to $R_S$ to $200R_S$ and the calculation is close to $1.0c$ the same as previous cases.

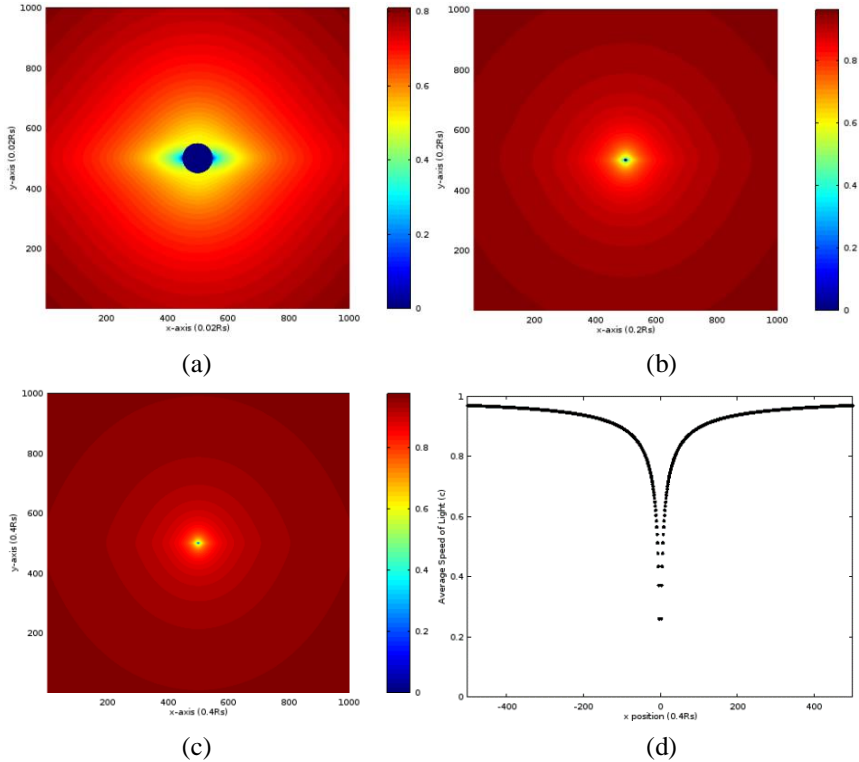

(a)      (b)

(c)      (d)

Figure 1. Distribution of the average radial speed of light in units of $c$, calculated from $r=R_S$ to the random point S, in the case that $a=1.0R_S$ and $R_Q=0.1R_S$. The maximums of x and y are both (a) $10R_S$, (b) $100R_S$, and (c) $200R_S$. (d) The distribution of the average radial speed of light along the x-axis in (c).

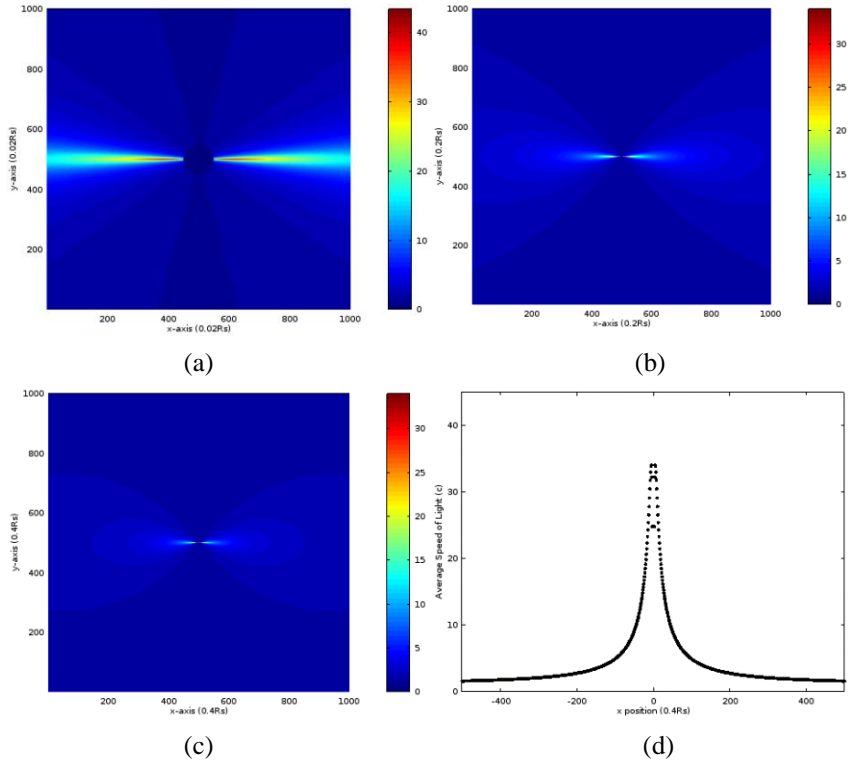

Figure 2. Distribution of the average radial speed of light in units of $c$, calculated from $r=R_S$ to the random point $S$, in the case that $a=100R_S$ and $R_Q=0.1R_S$. The maximums of x and y are both (a) $10R_S$, (b) $100R_S$, and (c) $200R_S$. (d) The distribution of the average radial speed of light along the x-axis in (c).

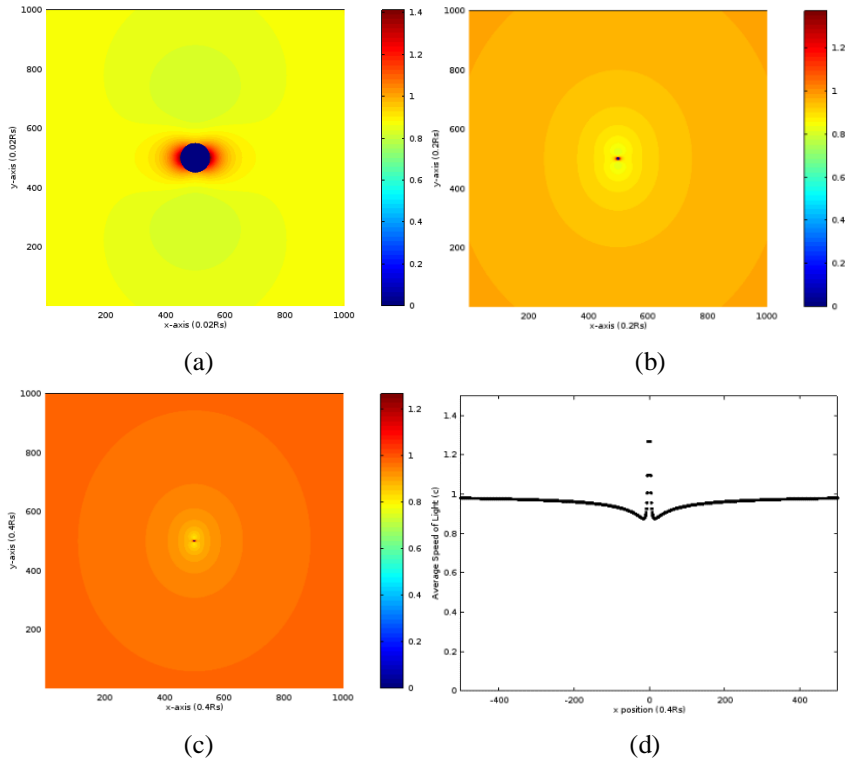

Figure 3. Distribution of the average radial speed in units of $c$, calculated from $r=R_S$ to the random point $S$, in the case that $a=1.0R_S$ and $R_Q=1.0R_S$. The maximums of x and y are both (a) $10R_S$, (b) $100R_S$, and (c) $200R_S$. (d) The distribution of the average radial speed of light along the x-axis in (c).

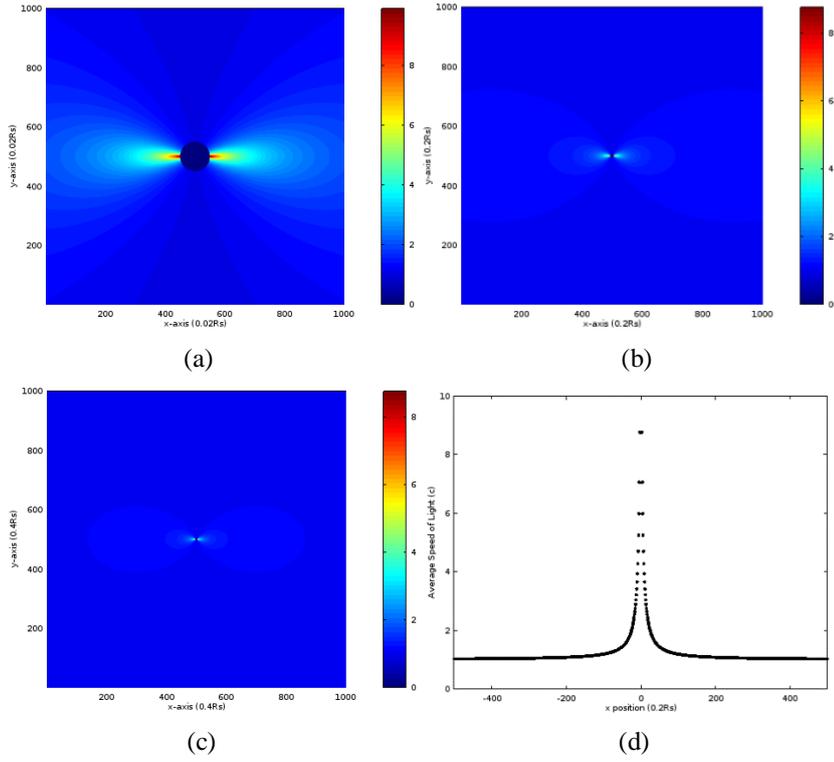

Figure 4. Distribution of average radial speed in units of *c*, calculated from $r=R_S$ to the random point *S*, in the case that $a=10R_S$ and $R_Q=1.0R_S$. The maximums of x and y are both (a) $10R_S$, (b) $100R_S$, and (c) $200R_S$. The distribution of the average radial speed of light in (c).

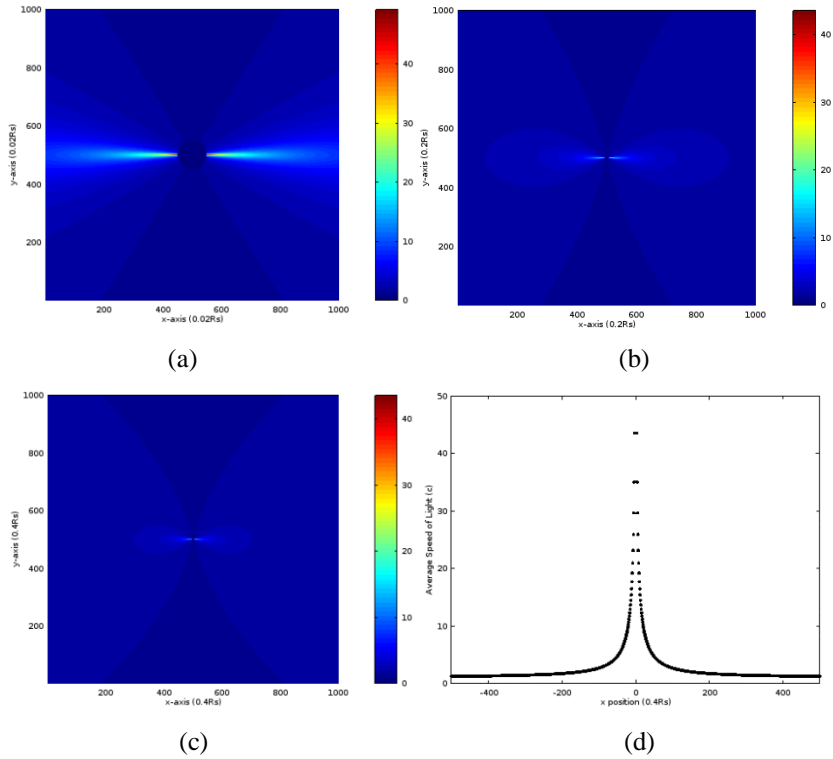

Figure 5. Distribution of the average radial speed of light in units of *c*, calculated from $r=R_S$ to the random point *S*, in the case that $a=50R_S$ and $R_Q=1.0R_S$. The maximums of x and y are both (a) $10R_S$, (b) $100R_S$, and (c) $200R_S$. The distribution of the average radial speed of light along the x-axis in (c).

The sixth case continues previous discussion where $R_Q$ is still $1.0R_S$ and $a$ is increased to $100R_S$. In Fig. 6(a), the maximum about $100.0c$ is also adjacent to $R_S$ on the x-axis. The distribution of the average radial speed more than $40.0c$ is much close to the x-axis symmetric to the center and its range is within $4.0R_S$. When enlarging the calculation space, it shows the average radial speed more than $40.0c$ close to the center as shown in Figs. 6(b) and (c). The distribution along the x-axis is shown in Fig. 6(d) where it quickly drops to $1.0c$ from $R_S$ to $200R_S$ and the calculation reveals the same result as the previous cases.

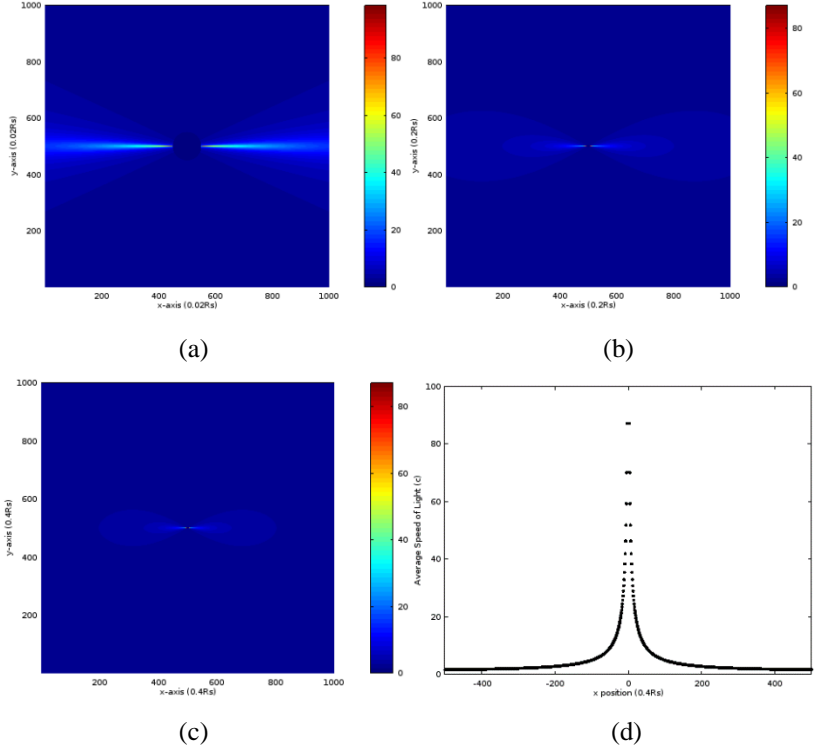

(a) (b)

(c) (d)

Figure 6. Distribution of the average radial speed of light in units of $c$, calculated from $r=R_S$ to the random point S, in the case that $a=100R_S$ and $R_Q=1.0R_S$. The maximums of x and y are both (a) $10R_S$, (b) $100R_S$, and (c) $200R_S$. The distribution of the average radial speed of light along the x-axis in (c).

Then we compare these cases with each other and find out the effects of $R_Q$ and $a$. Each data is discrete and we connect them in a smooth curve. In Fig. 7(a), we investigate the effect of $R_Q$ using the calculations in Figs. 1(d), 2(d). 3(d), and 6(d). The figure shows $\log_{10}$ of all curves. Comparing Fig. 1(d) with 3(d), the case of $0.1R_Q$ shows the minimum adjacent to $R_S$, and the case of $1.0R_Q$ has maximum adjacent to $R_S$ and then drops to a minimum a little away from $R_S$. Both cases show very close values when $r$ is more than $80R_S$ and then gradually increases to zero when $r=200R_S$. For another two cases of $0.1R_Q$ and $1.0R_Q$ with the same $a=100R_S$, the smaller $R_Q$ shows a drop when $r$ is close to $R_S$ and the maximum is at the distance a little away from $R_S$, whereas the bigger one has a maximum adjacent to $Rs$. Both values almost overlap when $r$ is about larger than $4.0R_S$ and gradually decreases to zero as $r$ increases largely. So the larger rotational term $a$ needs longer distance to reach the average radial speed close to $1.0c$.

The same result also reveals in the four different rotation cases where $R_Q$ holds at $1.0R_S$ as shown in Fig. 7(b). All curves show the maximums adjacent to $R_S$ and decrease to zero as $r$ is large enough. It means when the observer is far away the black hole, the measurement of the average radial speed of light is close to $c$ as we measure on Earth. This result can be applicable for other superstar that produce strong gravity with high rotation.

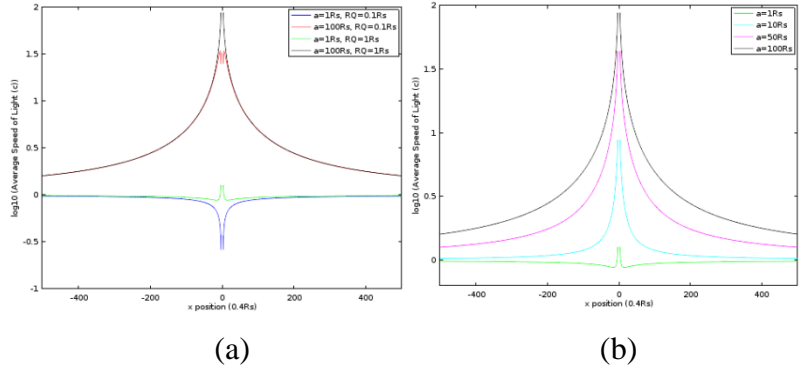

(a)          (b)

Figure 7. (a) $Log_{10}$ of the smooth curves from Figs. 1(d), 2(d), 3(d) and 6(d). (b) $Log_{10}$ of the smooth curves from Figs. 3(d) – 6(d).

## 6. Why the Speed of Massive Particle is astronomically Measured as Faster Than Light?

According to the previous results, the average radial speed of light is possible more than $c$ very tiny from the black hole to a far away place such as Earth as long as the radial speed is the dominant part. Some reports [1-7] showed that the speed of the particle away from the black hole was measured faster than $c$. This kind of phenomenon violates the special relativistic theory and makes us think about other possibility. Here we use our results to give another explanation.

As we know, there exists significant light bending near the superstar or black hole due to the strong gravity. The gravitation lensing is an example [23] Then we consider the relativistically electric particles leaving the black hole and radiating electromagnetic waves at two places A and B as shown in Fig. 8. Supposing the time difference measured in the Earth system is one year and the speeds of the particles are all less than $c$. So how can the measurements on Earth give the particles faster than $c$? As shown in Fig. 8, light radiated at the place A forwards to Earth and will be received after time $t$. Then when the particles move to the place B, light is radiated and will be received after time $t$-1 year on Earth, the actual path for the particles is along AB and the actual length is AB. Light emitted at A is along the trajectory with the average speed $c_1$ and light emitted at B is along the other one with the average speed $c_2$. The time difference between these two trajectories is 1 year and the former spending time is more than the later. Due to the strong gravitation, two trajectories are curves near the black hole and approximate two straight lines far away the black hole. The curvature of the light trajectory from A can be larger than that from B. Due to the light bending, the observed path of the particles is along A'B', not AB. We call the path A'B' the imaging path and its distance the imaging length. The phenomenon is much similar to the observation of particle in water by an observer above water. The different refraction indices between water and air cause light change its direction at the interface and results in the observer thinking the particle in shallow water. So the light bending near the black hole results in the imaging length A'B' longer than the actual length AB. The ratio of A'B' to AB might be several times so it gives the results of the particles faster than $c$.

After all, the special relativity [24] tells us that the massive particle cannot be speeded higher than $c$ and its relativistic mass close to infinite when the speed is very close to $c$. This principle has been verified in each operation on the synchrotron accelerator and it always needs a lot of energy to speed an electron close to $c$. We would expect that it is also the same phenomenon in the most places even near the black hole because the massive particle far away from the black hole with only finite total energy and its energy would be conserved even when it moves to the neighboring place of the black hole. However, when the speed of the massive particle is possible faster than the speed of light, it might be expected that the phenomenon like the Cherenkov's radiation would be observed in the regions near the black hole, some supermassive stars, or planets with strong gravity.

The speed of light near massively rotating and charged black hole can be faster than it on earth. This result can be one way to explain the flares near the black hole at IC 310 recorded on 12/13 November 2012 [25].

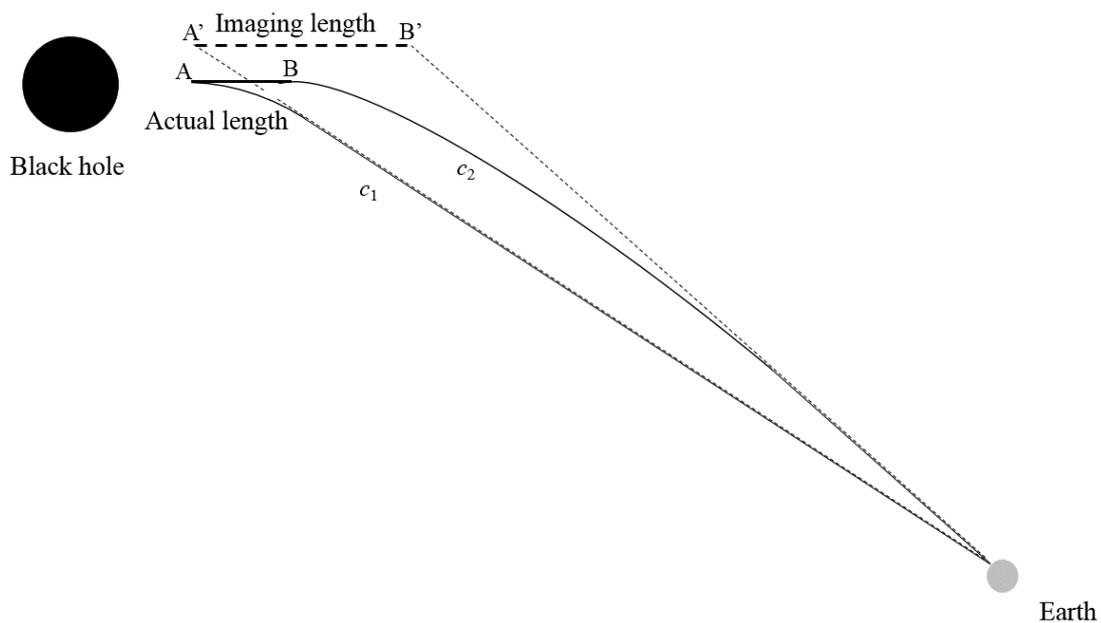

Figure 8. Two light trajectories with different average speeds cause the visual illusion in astronomical observations. It would result in the conclusion of the faster-than-light particle. Here $c_1$ and $c_2$ is very close to $c$ and possibly more or less than $c$ very tiny.

## 7. Is The Propagation Speed of Gravity Faster Than Light?

We might ask whether the propagation speed of gravity is faster than light or not. Considering a black hole with a spherical event horizon for convenient discussion and the result is applicable for other black hole with a non-spherical event horizon. If the black hole absorbs some mass in early time, then the event horizon will expand to a new event horizon with increase radius $\Delta r$ as shown in Fig. 9. When the propagation speed of gravity is faster than light, then the observation of a particle entering into the event horizon before expansion will be later than another observation of other particle entering into the new event horizon. It seems to cause an unreasonable phenomenon that the former particle enters into the black hole will be later than the later particle dose. Traditional theory thinks massive and massless particles cannot escape the black hole. The propagation speed of gravity faster than the measured speed of light will result in the disappear of any particle before reaching the event horizon, and the new event horizon covers all information of the particles passing through the former event horizon. This induces another problem that the event horizon is un-defined. So the propagation speed of gravity should not be faster than the measured speed of light.

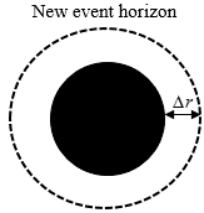

Figure 9. A spherical event horizon of a black hole expands to a new event horizon with increase radius $\Delta r$.

## 8. Conclusion

We use the Kerr-Newman metric based on General Relativity to discuss the average radial speed of light at the black hole. First, according to equivalence principle, time dilation requires some conditions between $R_S$, $a$, and $R_Q$. The appropriate range of $r$ is also given when light is along the radial direction. The Kerr-Newman metric shall exist everywhere so the concept of finite-size nucleus is better than a singularity at the center of the black hole and the totally enclosed charge $Q$ should be a function of $r$. The geodesic of light is determined by $ds^2=0$ then we obtain the velocity equation of light described on the reference frame far away from the black hole like on Earth. It is also possible described at the center of the black hole when it has finite-size nucleus. Next, we can calculate the spending time for light traveling from $r=\alpha_1 R_S$ to $r=\alpha_2 R_S$ at any $\theta$. We find that the average speed of light along the radial direction is possible larger than $c$ dependent on $a$ and $R_Q$. The larger $a$ is, the higher average radial speed of light is. The larger $R_Q$ also benefits higher average radial speed. Higher rotation or larger charge of the black hole gives a longer range where the average radial speed is more than *c* and it needs longer distance to decrease and close *c*. When an observer is far away from the black hole or other strong gravity, the measured average speed of light is close to *c* as the measurement on Earth. The results are reasonably at least for the radial directions from two poles and the place in the equatorial plane. According to the principle of equivalence, the accelerating directions are along the radial directions.

Based on these results, we propose a new explanation for the observation of the faster-than-light massive particles in astronomy. The light bending near the black hole or supermassive star with very strong gravity results in the arriving light beams having time difference by the telescope. The two events are recorded with a visual angle and it is thought that the massive particle moves a longer distance than reality. The phenomenon is like the observation above water that the positions of things in water are actually deeper than what we see.

Finally, we consider the increase of event horizon due to the expansion of the black hole to analyze whether the propagation speed of gravity can be faster than the measured speed of light or not. If it were true, then the event of a particle entering the new event horizon will be observed faster than the event of a particle entering the former event horizon. The traditional theory of the black hole thinks massive and massless particles cannot escape the black hole, and the phenomenon tells us that the former information is covered by the new event horizon and the former particle disappears before the event horizon. It seems to be unreasonable so the propagation of gravity should be slower than the measured speed of light.

## Acknowledgement

This research is under no funding.

**Reference:**
[1]. I. F. Mirabel and L. F. Rodríguez, *Nat*. **371**, 46 (1994).
[2]. R. D. Blandford, C. F. Mckee, and M. J. Rees, *Nat*. **267**, 211 (1977).
[3]. L. F. Rodríguez and I. F. Mirabel, *Natl*. *Acad*. *Sci*. *USA* **92**, 11390 (1995).
[4]. T. Belloni, M. Méndez, A. R. King, M. Van Der Klis, and J. Van Paradijs, *Astrophys*. J. **479**, L145 (1997).
[5]. Jerome A. Orosz, Erik Kuulkers, Michiel van der Klis, Jeffrey E. McClintock, Michael R. Garcia, Paul J. Callanan, Charles D. Bailyn, Raj K. Jain, and Ronald A. Remillard, *Astrophys*. *J*. **555**, 489 (2001).
[6]. C. C. Cheung D. E. Harris, and Ł. Stawarz, *Astrophys*. *J*. **663**, L65 (2007).
[7]. Keiichi Asada, Masanori Nakamura, Akihiro Doi, Hiroshi Nagai, and Makoto Inoue, *Astrophys*. *J*. *Lett*. **781**, L2 (2014).
[8]. E. T. Newman, R. Couch, K. Chinnapared, A. Elton, A. Prakash, and R Torrence, *J*. *Math*. *Phys*. **6**, 918 (1965).
[9]. Dani C. Wilkins, *Phys*. *Rev*. *D* **5**, 814 (1972).
[10]. Tim Adamo and E. T. Newman, arXiv: 1410.6626 (2016).
[11]. Irwin I. Shapiro, *Phys*. *Rev*. *Lett*. **13**, 789 (1964).
[12]. Irwin I. Shapiro, Michael E. Ash, Richard P. Ingalls, William B. Smith, Donald B. Campbell, Rolf B. Dyce, Raymond F. Jurgens, and Gordon H. Pettengill, *Phys*. *Rev*.

*Lett*. **26**, 1132 (1971).

[13]. Tomislav Kundi'c, Edwin L. Turner, Wesley N. Colley, J. Richard Gott, James E. Rhoads, Yun Wang, Louis E. Bergeron, Karen A. Gloria, Daniel C. Long, Sangeeta Malhotra, *Astrophys. J.* **482**, 75 (1997).

[14]. J. E. J. Lovell, D. L. Jauncey, J. E. Reynolds, M. H. Wieringa, E. A. King, A. K. Tzioumis, P. M. McCulloch, and P. G. Edwards, *Astrophys. J.* **508**, L51 (1998).

[15]. A. D. Biggs, I. W. A. Browne, P. Helbig, L. V. E. Koopmans, P. N. Wilkinson and R. A. Perley, *Mon. Not. R. Astron. Soc.* **304**, 349 (1999).

[16]. P. B. Demorest, T. Pennucci, S. M. Ransom, M. S. E. Roberts, and J. W. T. Hessel, *Nat.* **467**, 1081 (2010).

[17]. G. Risaliti, F. A. Harrison, K. K. Madsen, D. J. Walton, S. E. Boggs, F. E. Christensen, W. W. Craig, G. W. Grefenstette, C. J. Hailey, E. Nardini, Daniel Stein, and W. W. Zhang, *Nat.* **494**, 449 (2013).

[18]. F. De Felice and C. J. S. Clarke, *Relativity On Curved Manifolds* (Cambridge University Press, Cambridge, 1990), p. 355.

[19]. Bernard F. Schutz, *A First Course In General Relativity* (Cambridge University Press, Cambridge, 1985), p.291.

[20]. Hans C. Ohanian and Remo Ruffini, *Gravitation and Spacetime* (W. W. Norton & Company, 2$^{nd}$ ed., New York, 1994), p.445.

[21]. K. A. Pounds, C. J. Nixon, A. Lobban, and A. R. King, Monthly Notice Of The Royal Astronomical Society **481**, 1832 (2018).

[22]. I. S. Gradshteyn and I. M. Ryzhik, *Table Of Integrals, Series, And Products* (Academic Press, Burlington, 7$^{th}$ Ed., 2007), P. 254.

[23]. Jürgen Ehlers and Wolfgang Rindler, *Gen. Rel. Gravit.* **29**, 519 (1997).

[24]. Jerry B. Marion and Stephe T. Thornton, *Classical Dynamics of Particles & Systems* (Harcourt Brace Jovanovich, Inc., Orlando, 3$^{rd}$ ed., 1988), P. 507.

[25]. J. Aleksić et al., *Sci.* **346**, 1080 (2014).